\documentclass[twocolumn]{openjournal}
\usepackage{graphicx} 

\usepackage[colorlinks,linkcolor=blue,citecolor=blue,urlcolor=blue ]{hyperref}
\usepackage[utf8]{inputenc}
\usepackage{float}
\usepackage{xcolor}
\usepackage{ulem}
\usepackage{hyperref}	
\usepackage[T1]{fontenc}
\usepackage{amsmath}
\usepackage{xparse}
\usepackage{natbib} 
\usepackage{physics}
\usepackage{multirow}

\newcommand{\msolar}{\rm{M}_\odot}
\newcommand{\enzo}{\texttt{Enzo~}}
\newcommand{\swift}{\texttt{SWIFT~}}
\newcommand{\gadget}{\texttt{}{Gadget-2~}}

\begin{document}
\title{Halo Mass Functions at High Redshift\vspace{-1.5cm}}
\author{Hannah O'Brennan$^{1,*}$}
\author{John A. Regan$^1$}
\author{Chris Power$^{2,3}$}
\author{Saoirse Ward$^{1}$}
\author{John Brennan$^{1}$}
\author{Joe McCaffrey$^{1}$}
\thanks{$^*$E-mail:hannah.obrennan.2021@mumail.ie}
\affiliation{$^1$Centre for Astrophysics and Space Science Maynooth, Department of Physics, Maynooth University, Maynooth, Ireland \\
$^2$International Centre for Radio Astronomy Research (ICRAR), M468, University of Western Australia, 35 Stirling Hwy, Crawley,
WA 6009, Australia\\
$^3$ARC Centre of Excellence for All Sky Astrophysics in 3 Dimensions (ASTRO 3D)}

\begin{abstract}
\noindent Recent JWST observations of very early galaxies, at $\rm{z \gtrsim 10}$,  have led to claims that tension exists between the sizes and luminosities of high-redshift galaxies and what is predicted by standard $\Lambda$CDM models. Here we use the adaptive mesh refinement code \enzo and the N-body smoothed particle hydrodynamics code \swift to compare (semi-)analytic halo mass functions against the results of direct N-body models at high redshift. In particular, our goal is to investigate the variance between standard halo mass functions derived from (semi-)analytic formulations and N-body calculations and to determine what role any discrepancy may play in driving tensions between observations and theory. We find that the difference between direct N-body calculations and (semi-) analytic halo mass function fits is less than a factor of 2 (at $\rm{z \sim 10}$) within the mass range of galaxies currently being observed by JWST, and is therefore not a dominant source of error when comparing theory and observation at high redshift.  
\end{abstract}

\maketitle

\section{Introduction} \label{Sec:Introduction}
\noindent Since its launch in December 2021 and its subsequent data releases in the months since, JWST has both reshaped and challenged our understanding of galaxy formation in the very early Universe ($z \gtrsim 10$). In particular, JWST has discovered a number of very massive and very luminous galaxies at redshifts in excess of $z = 10$ which, after initial photometric detection, have now been spectroscopically confirmed \citep{Harikane_2024, ArrabalHaro_2023, Castellano_2024, Hainline_2024, Gentile_2024}. 
These galaxies, both in terms of their intrinsic luminosity and their potential host halo masses, provide a significant challenge to our understanding of structure formation in the early Universe. \cite{Boylan-Kolchin_2023}, using an analytic model applied to the inferred JWST stellar masses of a number of high-redshift sources from \cite{Labbe_2023}, found that the stellar masses implied by the sources required star formation efficiencies, $\epsilon$, significantly in excess of those from the present-day universe and perhaps as implausibly high as $\epsilon = 1$. The model employed by 
\cite{Boylan-Kolchin_2023} hinges on a number of simple yet strong assumptions. These assumptions use the inferred stellar host mass to derive a host halo mass based on the ratios between the cosmic baryon density and the cosmic matter density and the star formation efficiency. 
These simple arguments culminate in a calculation of the probability of finding such (luminous) galaxies at early times in a $\Lambda$CDM universe. \\
\indent \cite{Boylan-Kolchin_2023} conclude that the most massive JWST galaxies detected are both at the very limit of galaxy formation theory and that their number densities are difficult to equate with the JWST field of view. They conclude that these issues indicate that there are several unresolved issues in our theories. However, underneath the model employed by \cite{Boylan-Kolchin_2023} are a number of astrophysical assumptions which are used when converting the broadband spectral energy distribution to a host stellar mass. This conversion is open to large uncertainties - particularly since our knowledge of how to do this conversion comes from the local Universe. There is no guarantee that these conversion relations can be directly mapped to the high-$z$ Universe and in fact most analysis shows that it is almost certainly not the case \citep[e.g.][]{Kannan_2023, Lu_2024}. For example, recent analysis by \cite{Steinhardt_2023} show that modifying the host population IMF of the inferred stellar population results in a decrease in the stellar mass by factors of between 10 and 50. Such decreases in the host stellar mass have a direct knock-on effect to the inferred host halo mass and can significantly decrease the tension with  $\Lambda$CDM models.\\
\indent Adding to this important point is that \cite{Boylan-Kolchin_2023} (as well as many other studies) utilise the well-tested and parameterised halo mass function (HMF) derived by \cite{Sheth_1999} in order to compute their galaxy number densities and cumulative comoving number density of galaxies. An important consideration therefore is to test, using explicit N-body calculations, how accurate the underlying HMF is at high-$z$ (i.e. at $z \gtrsim 10$), compared to direct N-body calculations, and what error may be associated with this model. This is the goal of this study. \\
\indent While preparing this work a study by \cite{Yung_2024} performed a similar investigation. Using a high-fidelity suite of N-body simulations across a broad range of box sizes and redshifts, \cite{Yung_2024} were able to show that HMFs derived from fitting functions and analytical approximations match extremely well to their N-body simulations. \cite{Yung_2024} found that even up to $z = 15$ the match between their N-body simulations and the fitting functions is no more than approximately a factor of two across a range of fitting functions. As discussed above, providing robust quantification of the match between fitting functions and direct N-body calculations at high-$z$ is extremely timely given the recent JWST results. \\
\indent This is particularly relevant when trying to understand the probability of 
finding such luminous and massive galaxies within a JWST field of view. 
Thus far several studies have tested the first results from JWST against state-of-the-art hydrodynamical simulations and the results agree within a factor of a few \citep[e.g.][]{Keller_2023, McCaffrey_2023, Sun_2023, Rennehan_2024}. It is therefore timely to quantify potential sources of systematic error when comparing observations and models.  \\
\indent Here we perform a similar analysis to \cite{Yung_2024}, with the difference being that we compare HMF fitting functions against two fundamentally different numerical codes - \enzo \citep{bryan_enzo_2014, BrummelSmith_2019} and \swift \citep{schaller_swift_2018, schaller_swift_2024}. While \cite{Yung_2024} used the publicly available \gadget code (\citet{Springel_2005}) which is a TreePM code, with a similar gravitational solver to \swift, we also use the Particle-Mesh based \enzo code which gives an additional layer of comparison. Our analysis confirms the results of \cite{Yung_2024} and we also observe a factor of approximately two difference between the HMFs generated by the N-body codes and the fitting functions. \\
\indent The structure of the paper is as follows: In \S \ref{Sec:Methodology} we outline the methodology including the simulations and fitting functions employed. In \S \ref{Sec:Results} we deliver the results of our analysis and in \S \ref{Sec:Discussion} we summarize and discuss our results in light of recent JWST observations.

\section{Methodology} \label{Sec:Methodology}
\subsection{Numerical simulations}
\noindent We run a series of dark matter-only simulations using \enzo and \swift of varying resolutions, achieved by varying both the box size and particle number.
The box size varies from $L=$ 0.5 cMpc/h to 100.0 cMpc/h and the particle number varies from $N_{\text{DM}}=$ 512$^{3}$ to 1024$^{3}$. 
Details of the simulation boxes are summarised in Tables \ref{tab:mass_res} and \ref{tab:space_res}.

\begin{table}
    \centering
    \begin{tabular}{|c|c|c|}
         \hline
         $L$ [cMpc/h]&  $N_{\text{DM}}^{1/3}$ &  $ \rm{M_{\text{DM}}}$ [M$_{\odot}$/h]\\
         \hline
         0.5 &  512 &  6.69 $\times$ 10$^{1}$ \\
         1.5 &  512 &  1.81 $\times$ 10$^{3}$ \\
         2.5 &  512 &  8.37 $\times$ 10$^{3}$ \\
         7.5 &  1024 &  2.83 $\times$ 10$^{4}$ \\
         7.5 &  512 &   2.26 $\times$ 10$^{5}$ \\
         12.5 &  1024 &  1.31 $\times$ 10$^{5}$ \\
         12.5 &  512 &  1.05 $\times$ 10$^{6}$ \\
         25.0 &  1024 &  1.05 $\times$ 10$^{6}$ \\
         25.0 &  512 &  8.37 $\times$ 10$^{6}$ \\
         50.0 &  1024 &  8.37 $\times$ 10$^{6}$ \\
         50.0 &  512 &  6.70 $\times$ 10$^{7}$ \\
         100.0 &  1024 &  6.70 $\times$ 10$^{7}$ \\
         100.0 &  512 &  5.36 $\times$ 10$^{8}$ \\
         \hline
    \end{tabular}
    \caption{Mass resolutions of each simulation box. All simulations but the $L=7.5$ cMpc/h, $N_{\text{DM}}^{1/3} = 1024$ box are run using both \enzo and \swift; this box is run using \swift only.}
    \label{tab:mass_res}
\end{table}

\begin{table}
    \centering
    \begin{tabular}{|c|c|c|c|}
         \hline
         $L$ [cMpc/h]&  $N_{\text{DM}}^{1/3}$ &  $ \Delta x_{\text{Enzo}}$ [cpc/h] &  $ \Delta x_{\text{SWIFT}}$ [cpc/h]\\
         \hline
         0.5 & 512 & 7.63 $\times$ 10$^{0}$ & 3.91 $\times$ 10$^{1}$ \\
         1.5 & 512 & 2.29 $\times$ 10$^{1}$ & 1.17 $\times$ 10$^{2}$ \\
         2.5 & 512 & 3.82 $\times$ 10$^{1}$ & 1.95 $\times$ 10$^{2}$ \\
         7.5 & 1024 & - & 2.93 $\times$ 10$^{2}$ \\
         7.5 & 512 & 1.14 $\times$ 10$^{2}$ & 5.86 $\times$ 10$^{2}$ \\
         12.5 & 1024 & 9.54 $\times$ 10$^{1}$ & 4.88 $\times$ 10$^{2}$ \\
         12.5 & 512 & 1.91 $\times$ 10$^{2}$ & 9.77 $\times$ 10$^{2}$ \\
         25.0 & 1024 & 1.91 $\times$ 10$^{2}$ & 9.77 $\times$ 10$^{2}$ \\
         25.0 & 512 & 3.82 $\times$ 10$^{2}$ & 1.95 $\times$ 10$^{3}$ \\
         50.0 & 1024 & 3.82 $\times$ 10$^{2}$ & 1.95 $\times$ 10$^{3}$ \\
         50.0 & 512 & 7.63 $\times$ 10$^{2}$ & 3.91 $\times$ 10$^{3}$ \\
         100.0 & 1024 & 7.63 $\times$ 10$^{2}$ & 3.91 $\times$ 10$^{3}$ \\
         100.0 & 512 & 1.53 $\times$ 10$^{3}$ & 7.81 $\times$ 10$^{3}$ \\
         \hline
    \end{tabular}
    \caption{Highest space resolutions of each simulation box. All simulations but the $L=7.5$ cMpc/h, $N_{\text{DM}}^{1/3} = 1024$ box are run using both \enzo and \swift; this box is run using \swift only.}
    \label{tab:space_res}
\end{table}

Our simulations begin at $z=127.0$ with initial conditions set using MUSIC (\citet{hahn_multi-scale_2011}) and end at $z=10.0$. 
We use a $\Lambda$CDM cosmology with $\text{h}=0.6774$, $\Omega_{\text{m}, 0}=0.2592$, $\sigma_{8}=0.8159$,
$n_{\text{eff}}=0.9667$ 
and the \citet{eisenstein_baryonic_1998} transfer function for no baryon acoustic oscillations. The two codes we use, \enzo  and \swift, are both well-tested and have been used extensively within the community. Additionally, both codes use somewhat different strategies for solving the Poisson equation which allows for additional comparison between the various semi-analytic fits against numerical solutions. We now describe both codes but refer the interested reader to the code method papers for more details. 
 
\enzo is a grid-based N-body code with the capability for adapative mesh refinement (AMR), widely used in cosmological hydrodynamics simulations.
The AMR allows for improved resolution in areas of interest (e.g. collapsing structures) without greatly increasing computational cost and without needing prior knowledge of the volume to pre-select areas for increased refinement. 
The gravity solver works by implementing a Fast Fourier Transform (FFT) technique to solve Poisson's equation at the root grid of each timestep of the simulation. The boundary conditions on the subgrids are then interpolated from the parent grid and the Poisson equation is solved at each time step, one subgrid at a time. 
For a unigrid \enzo simulation (i.e. no refinement), the minimum inter-particle separation in which gravity acts is twice the length of two cells, given as:

\begin{equation}
    \Delta x_{\text{Unigrid}} = 2\frac{L}{N_{\text{DM}}^{1/3}}.
\end{equation}

When running \enzo simulations, we used a maximum refinement level of 8, meaning that the resolution was increased by a factor 2$^{8}$ in regions of high particle density. This means the highest resolution is given as:

\begin{equation}\label{eqn:delta_x}
    \Delta x_{\text{Enzo}} = \frac{1}{2^{8}}\Delta x_{\text{Unigrid}} = \frac{1}{2^{7}}\frac{L}{N_{\text{DM}}^{1/3}}.
\end{equation}

For the dark matter-only simulations carried out here, refinement is triggered once the particle over-density reaches a factor of 4 greater than the mean density (i.e. \texttt{MinimumOverDensityForRefinement = 4}). In addition to this we set the \texttt{MinimumMassForRefinementLevelExponent = -0.1} which makes the refinement scheme super-Lagrangian and allows for the higher levels of refinement to be more easily triggered. The combination of these parameter choices mean that our simulation setup is somewhat conservative, employing an aggressive refinement strategy \citep[see e.g.][]{OShea_2005}. 

\textit{SPH With Inter-dependent Fine-grained Tasking}  - \swift - combines a tree-based N-body solver with a smoothed particle hydrodynamics (SPH) solver.  In this study, we use the adaptive mode of the Fast Multipole Method (FMM) (\citet{Cheng_1999}). This implements a Taylor expansion twice to resolve the gravitational potential (and later the forces) between particles in different cells. We set the accuracy criterion $\epsilon_{\text{FMM}}=0.001$. \swift takes advantage of the hierarchical tree structure to efficiently solve for the gravitational forces between particles. Particles from nearest neighbour cells are treated as individuals. A group of particles from distant cells are approximated as one particle with the total mass of the group located at the centre of mass. Long-range forces are resolved using a Fast Fourier Transform algorithm (\citet{Frigo_Johnson_2005}). \\
\indent It is a non-trivial matter comparing \enzo and \swift due to the difference in the gravity solvers. The highest spatial resolution for an \enzo simulation, $\Delta x_{\text{Enzo}}$, applies only to regions of high particle density. The softening length for a \swift simulation, $\Delta x_{\text{SWIFT}}$, applies to the entire simulation volume. We compromise by setting the softening length to an intermediate value, between $\Delta x_{\text{Unigrid}}$ and $\Delta x_{\text{Enzo}}$: 

\begin{equation}
    \Delta x_{\text{SWIFT}} = \frac{1}{5^{2}}\frac{L}{N_{\text{DM}}^{1/3}}.
\end{equation}

\subsection{Numerical Halo Finders}
As discussed in \S \ref{Sec:Introduction} the goal of this paper is to determine the differences between (semi-)analytic HMFs and those derived from direct N-body simulations. In order to determine the HMF from the 
cosmological simulations we must employ a halo finder and decide on the redshifts at which to evaluate the HMFs. \\
\indent We analyse the simulation snapshots from our full suite of outputs at $z=20.0$, $z=15.0$ and $z=10.0$ using a  friends-of-friends (FOF) (\citet{efstathiou_numerical_1985}) and HOP halo (\citet{eisenstein_hop_1998}) finder (results from the HOP finder can be found in the appendix of the paper as the results are very similar between FOF and HOP and the goal of this study is not to compare halo finders). The FOF halo finder accounts for the distances between dark matter particles within a single snapshot. We use a linking length $\Delta x(b) = b \times L/N_{\text{DM}}^{1/3}$, where $L/N_{\text{DM}}^{1/3}$ is the mean inter-particle separation. This refers to the maximum permitted separation between two particles. 
A group i.e. halo is found from a set of inter-linked particles. 

Our halos are approximated as spheres and ideally, we choose $b$ such that each halo encompasses a volume with an overdensity $\Delta_{\text{c}} \approx 18\pi^{2}$ i.e. 178 times the critical density $\rho_{\text{c}}(z)$ (\citet{Bryan_1998}).
The Python package \texttt{hmf} \citep{murray_hmfcalc_2013} provides an approximation relating $\Delta_{\text{c}}$ and $b$:

\begin{equation}
    \Delta_{\text{c}}(z) = \frac{9}{2\pi b^{3}} \, \Omega_{\text{m}}(z).
\end{equation}

As $z \rightarrow \infty$, $\Omega_{m}(z) \rightarrow 1$ in our cosmology. Therefore, $b=0.2$ results in a predicted overdensity of $\Delta_{\text{c}} \approx 178$ and this is the value we choose for our FOF algorithm. 

The HOP halo finder accounts for the distances between particles as well as the computed density of each particle. Rather than creating a continuous density field in the simulation volume, the density of each particle at its given position is estimated using a spherically symmetric cubic spline kernel (\citet{Monaghan_Lattanzio_1985}) on its $N_{\text{dens}} = 64$ nearest neighbours. The density is normalised to the average density of particles within the simulation box e.g. a particle with a computed density of $\delta=80$ refers to a position 80 times denser than the average i.e. an overdensity (\citet{Parallel_HOP}). A link is made by hopping from a given particle to the densest of its $N_{\text{hop}} = 64$ neighbours. This process continues, forming a chain of increasing particle density until we reach a particle that is its own densest neighbour.  All chains sharing the same densest particle are part of the same group.
If the maximum particle density within a group is above a set threshold overdensity $\delta_{\text{peak}}$, that group is defined as a halo. 
We use a threshold overdensity of $\delta_{\text{peak}} = 100$. 
This value was chosen to calibrate the HOP finder with the FOF finder so both halo finders found a similar halo number and mass range for the \enzo, $L=0.5$ cMpc/h, $N_{\text{DM}}=512^{3}$, $z=10$ snapshot.

The halo masses found from these halo catalogues are the FOF and HOP masses i.e. the mass of the halo is defined as the sum of the masses of the particles that make up the halos.
After we create halo catalogues based on these halo finders, we only include halos made up of at least $N_{\text{DM, min}}=100$ particles to reduce numerical over-counting (i.e. halos must be well resolved (by at least 100 particles) before we identify them as halos).

\begin{figure}[h]
    \centering
        \includegraphics[width=0.4\textwidth]{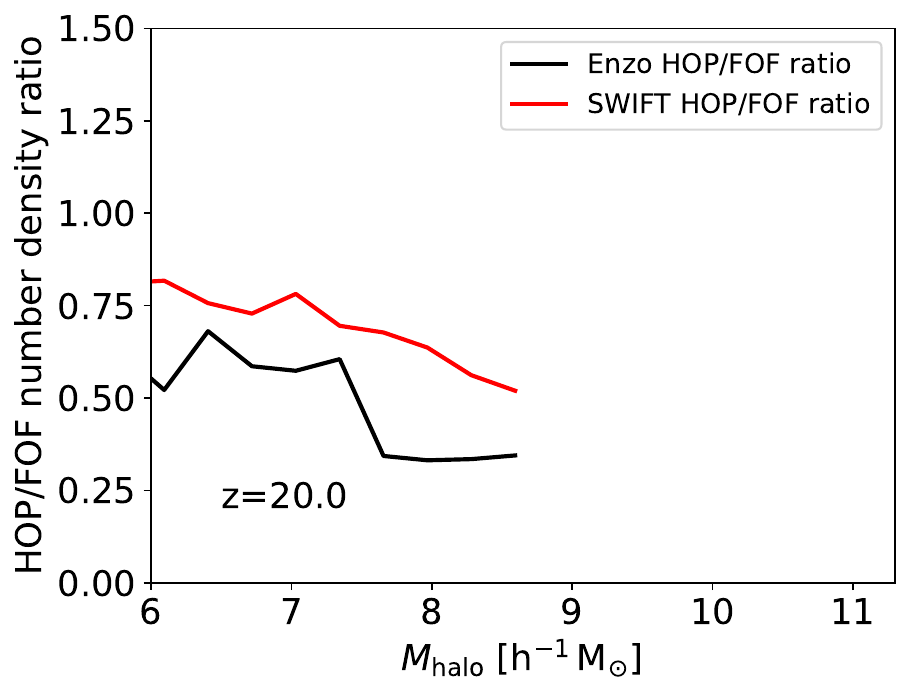}
    \hfill
        \includegraphics[width=0.4\textwidth]{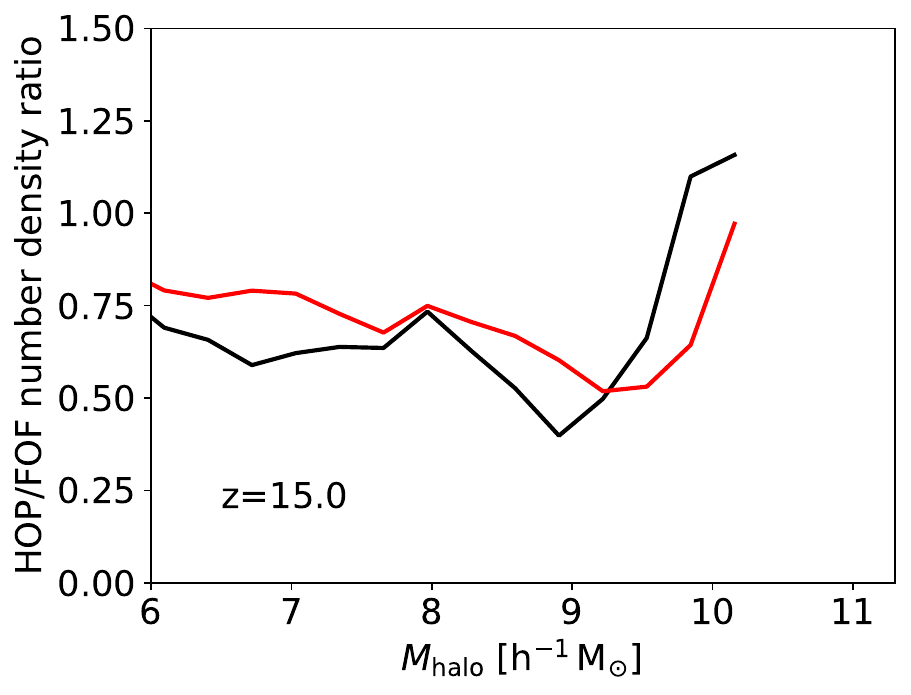}
    \hfill
        \includegraphics[width=0.4\textwidth]{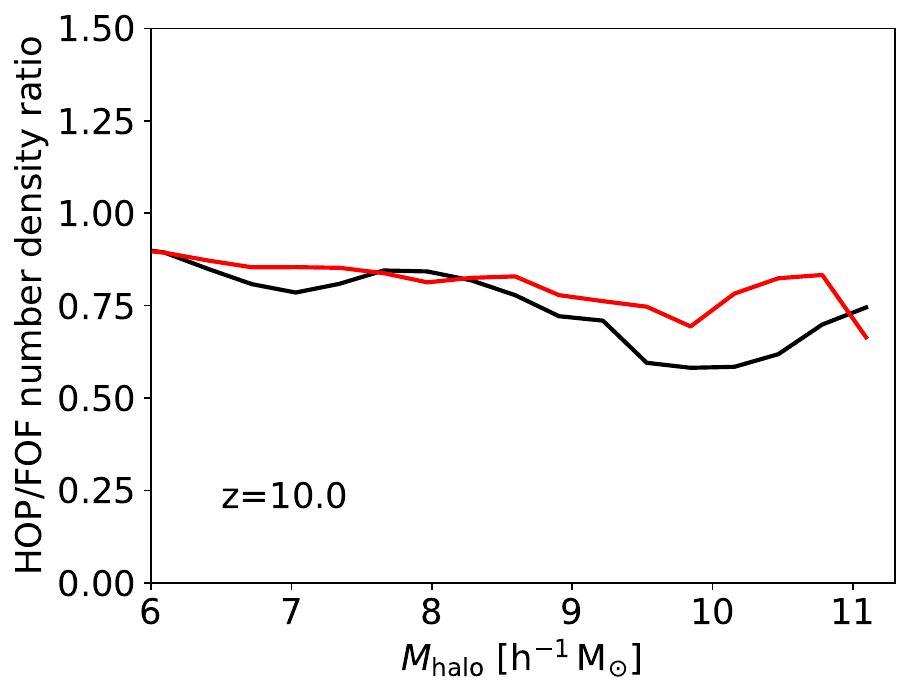}
     
\caption{Comparing the FOF and HOP halo number densities derived from \enzo and \swift simulation data (with halo masses on a logarithmic scale). We show how the ratio of the HOP number density to the FOF number density varies with halo mass. The HOP number density is within a factor of 2 of its FOF counterpart - particularly at the z = 10 outputs. Some larger deviations at higher z are seen as expected.}
\label{fig:FOF_HOP_comp}
\end{figure}

We find that there is little difference between the number densities found using FOF and HOP for a given simulation suite, with HOP underestimating the FOF number densities by a factor of $\approx$ 2 at most (see Figure \ref{fig:FOF_HOP_comp} for a comparison). Given that the halo finder parameters were calibrated at $z=10.0$, it is not surprising that there is little mass variance in their number density ratio at this redshift. From this point onwards we show the results for FOF only and leave those from HOP to the appendix.

\subsection{Numerical Halo Number Densities}
\noindent A halo catalogue is created from each snapshot with a specific halo finder using the \texttt{yt}\footnote{We use a version of \texttt{yt} called \texttt{yt-swift} developed by \citet{rennehan_yt-swift} as the most recently available version of \texttt{yt} cannot load \swift output data.} and \texttt{yt-astro-analysis} packages (\citet{turk_yt_2011}, \citet{smith_yt-projectyt_astro_analysis_2022}).
An array of halo mass values is generated from each halo catalogue, filtered to exclude halos consisting of $< \, N_{\text{DM, min}}$ particles. 
For each array, the halo mass values are binned on a logarithmic scale into 24 histogram bins with $M_{\text{min}} = 10^{3.75} \, \text{M}_{\odot}/\text{h}$ and $M_{\text{max}} = 10^{11.75} \, \text{M}_{\odot}/\text{h}$. 
Halo catalogues from a given simulation suite, halo finder and redshift, $z$, are combined, giving rise to 12 datasets (see Table \ref{tab:FOF_HOP_datasets} for more detail).
For each \enzo dataset, we combine halo catalogues based on 12 simulation boxes. For each \swift dataset, we combine halo catalogues based on 13 simulation boxes (see Tables \ref{tab:mass_res} and \ref{tab:space_res} for more detail).
In our algorithm below, we index datasets with $i \in \{1, \cdots, 12\}$, mass bins with $j \in \{1, \cdots, 24\}$, \enzo simulation boxes with $k \in \{1, \cdots, 12\}$, and \swift simulation boxes with $k \in \{1, \cdots, 13\}$.

For example, consider the $i=3$ dataset (\enzo, FOF, $z=10.0$). Say we wish to find the number density $n_{\text{halo, } i=3}^{\text{num}}(M_{\text{mid, } j})$ of the $j^{\text{th}}$ bin centred at some mass $\log_{10}(M_{\text{mid, } j}/\text{h}^{-1} \,\text{M}_{\odot})$. We count the number of halos $N_{\text{halo, } 3,j,k}$ in this bin for each simulation box $\forall k=1, \cdots, 12$ and this specific halo number density is given as:

\begin{equation}
    n_{\text{halo, } 3,j,k}^{\text{num}}(M_{\text{mid, } j}) = \frac{N_{\text{halo, } 3,j,k}}{V_{k}},
\end{equation}

where $V_{k}$ is the comoving volume of the $k^{\text{th}}$ simulation box.

We find the halo number density $n_{\text{halo, } 3}^{\text{num}}(M_{\text{mid, } j})$ by averaging over all non-zero number densities across all 12 simulation boxes:

\begin{equation}
    n_{\text{halo, } 3}^{\text{num}}(M_{\text{mid, } j}) = \frac{1}{S_{3,j}}\sum_{k}n_{\text{halo, } 3,j,k}^{\text{num}}(M_{\text{mid, } j}),
\end{equation}

where $S_{3,j}$ is the number of halo catalogues that found a non-zero halo number density for the $j^{\text{th}}$ bin in this dataset.

We can generalise the above equation for the $i^{\text{th}}$ dataset and $j^{\text{th}}$ bin as:

\begin{equation}
    n_{\text{halo, } i}^{\text{num}}(M_{\text{mid, } j}) = \frac{1}{S_{ij}}\sum_{k}n_{\text{halo, } ijk}^{\text{num}}(M_{\text{mid, } j}).
\end{equation}

\begin{table}
    \centering
    \begin{tabular}{|c|c|c|}
    \hline
        Simulation & Halo finder & Snapshot Redshift ($z$) \\
        \hline
        \multirow{6}{*}{\enzo} & \multirow{3}{*}{FOF} & 20.0 \\
         &  & 15.0 \\ 
         &  & 10.0 \\ \cline{2-3}
         &  \multirow{3}{*}{HOP} & 20.0 \\
         &  & 15.0 \\
         &  & 10.0 \\
         \hline\hline
         \multirow{6}{*}{\swift} & \multirow{3}{*}{FOF} & 20.0 \\
         &  & 15.0 \\
         &  & 10.0 \\ \cline{2-3}
         & \multirow{3}{*}{HOP} & 20.0 \\
         &  & 15.0 \\
         &  & 10.0 \\
         \hline\hline
    \end{tabular}
    \caption{The 12 datasets derived by combining halo catalogues for each simulation, halo finder, and redshift combination.}
    \label{tab:FOF_HOP_datasets}
\end{table}

\subsection{(Semi-)Analytical Halo Mass Functions}
We compare our numerical halo number densities with those derived from both analytical forms and popular fits.
The number density (units: $\text{h}^{3} \, \text{cMpc}^{-3}$) of dark matter halos at a given redshift, $z$, in a mass bin centred at $\log_{10}(M_{\text{mid}}/\text{h}^{-1} \, \text{M}_{\odot})$ with a width of $\Delta \log_{10}M$ is defined as:

\begin{equation} \label{eqn:n_halo_analytic}
\begin{split}
n_{\text{halo}}^{\text{fit}}(z, M_\text{mid}) &= \int_{M_{\text{a}}(M_{\text{mid}})}^{M_{\text{b}}(M_{\text{mid}})} \frac{dn}{dM}(z, M) \, dM, \\
\log_{10}M_{\text{a}}(M_{\text{mid}}) &= \log_{10}\left(\frac{M_{\text{mid}}}{\text{h}^{-1} \, \text{M}_{\odot}}\right) - \frac{\Delta \log_{10}M}{2}, \\
\log_{10}M_{\text{b}}(M_{\text{mid}}) &= \log_{10}\left(\frac{M_{\text{mid}}}{\text{h}^{-1} \, \text{M}_{\odot}}\right) + \frac{\Delta \log_{10}M}{2}.
\end{split}
\end{equation}

The halo mass function (units: $\text{h}^{4} \, \text{cMpc}^{-3} \, \text{M}_{\odot}^{-1}$) is the differential halo number density per unit mass and is defined as:

\begin{equation} \label{eqn:dndm_analytic}
\frac{dn}{dM}(z, M) = \frac{\rho_{0}}{M^{2}}\, f(\sigma(z, M)) \,\Big|\frac{d\ln\sigma(z, M)}{d\ln M} \Big|.    
\end{equation}

Here $\rho_{0}$ (units: $\text{h}^{2} \, \text{M}_{\odot} \, \text{cMpc}^{-3}$) and $\sigma(z, M)$ refer to the mean density of the universe and mass variance respectively. 
The mass variance is given as:

\begin{equation}
\sigma^{2}(z,M) = \frac{1}{2\pi^{2}}\int_{0}^{\infty} k^{2} \, P(z,k) \, \tilde{W}^{2}(kR(M)) \, dk,
\end{equation}

where $P(z, k)$ is the linear power spectrum and $\tilde{W}(kR)$ is a window function with a filter defined by $R = R(M)$. We use the Top Hat window function given as:

\begin{equation}
\tilde{W}(kR) = \frac{3}{(kR)^{3}}\left[\sin(kR) - (kR)\cos(kR)\right], 
\end{equation}

\begin{equation}
    W(r) =
        \begin{cases}
        \frac{3}{4\pi R^{3}} & \text{if } r < R\\
        0 & \text{if } r > R.
        \end{cases}
\end{equation}

The linear power spectrum evolves with $z$ as:

\begin{equation}
    P(z,k) = d(a(z))^{2} \, P(z=0,k), 
\end{equation}

where $a(z) = 1/(1+z)$ and $d(a)$ is the normalised linear growth factor (\citet{Lukic_2007}) given as:

\begin{equation}
d(a) = \frac{D^{+}(a)}{D^{+}(a=1)},
\end{equation}

\begin{equation}
D^{+}(a) = \frac{5\Omega_{m,0}}{2} \, \frac{H(a)}{H_{0}} \, \int_{0}^{a} \frac{da'}{\left[a'H(a')/H_{0}\right]^{3}} .
\end{equation}

The exact form of $P(z=0,k)$ depends on the cosmology and the choice of transfer function. Like with our simulations, we initialise the \texttt{hmf} objects with the \citet{eisenstein_baryonic_1998} transfer function with no baryon acoustic oscillations.
The exact form of $f(\sigma(z, M))$ depends on the choice of fitting function for the halo mass function.

\begin{figure}[h]
    \centering
        \includegraphics[width=0.4\textwidth]{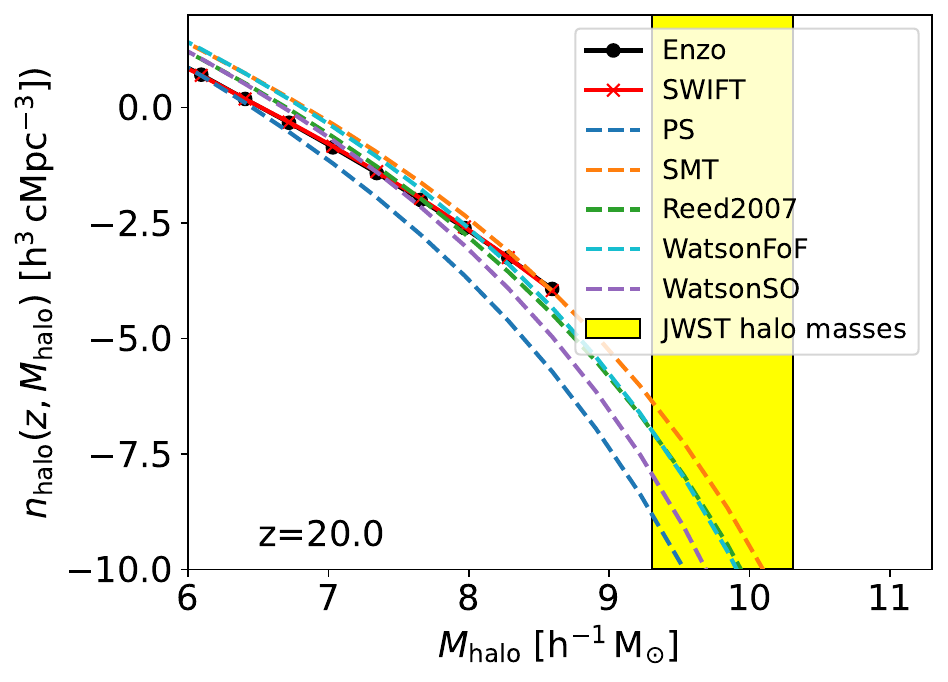}
    \hfill
        \includegraphics[width=0.4\textwidth]{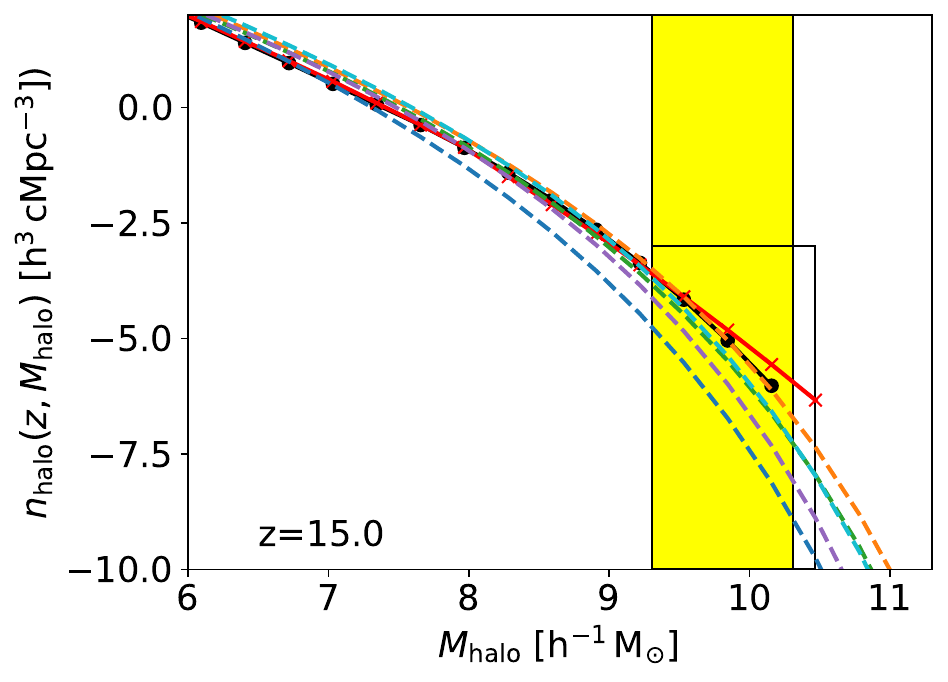}
    \hfill
        \includegraphics[width=0.4\textwidth]{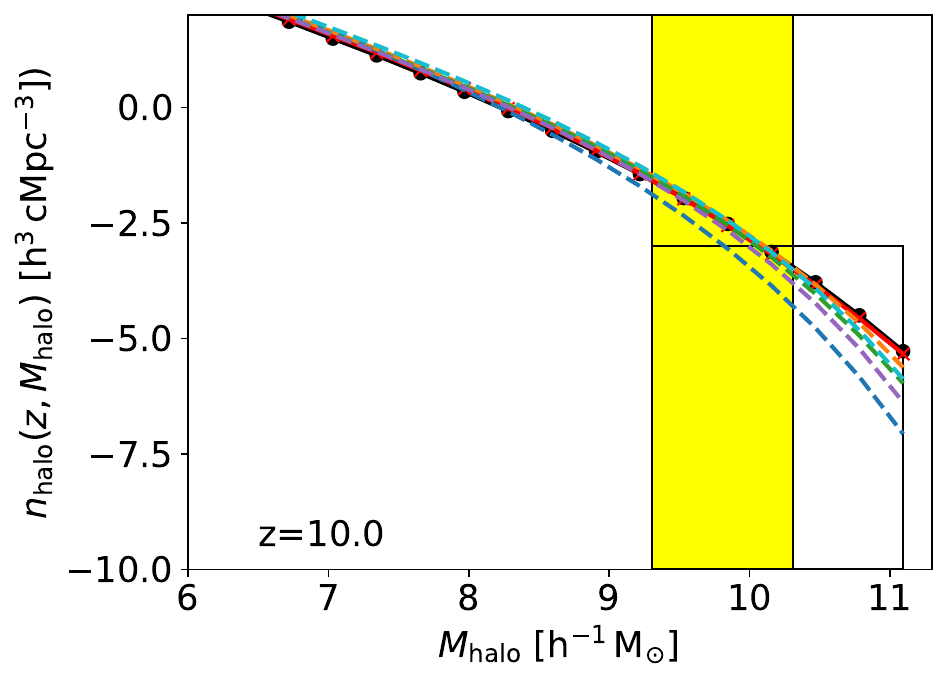}
     
\caption{Comparing the \enzo and \swift halo number densities with five halo number densities derived from fits (see Table \ref{tab:HMFs}) at $z=20.0$ (upper panel), $z=15.0$ (centre panel) and $z=10.0$ (lower panel) using the FOF halo finder (with halo masses and number densities on a logarithmic scale). The black rectangles indicate the regions for more detailed analysis as seen in Figure \ref{fig:FOFZoom}. The yellow shaded regions represent the approximate range of halo masses detected by JWST at $z \geq 10.0$. Over the halo mass range selected at $z=10.0$, all but the PS fit agree with numerical results within a factor of 2.}
\label{fig:FOF}
\end{figure}

\begin{figure}[h]
    \centering
        \includegraphics[width=0.4\textwidth]{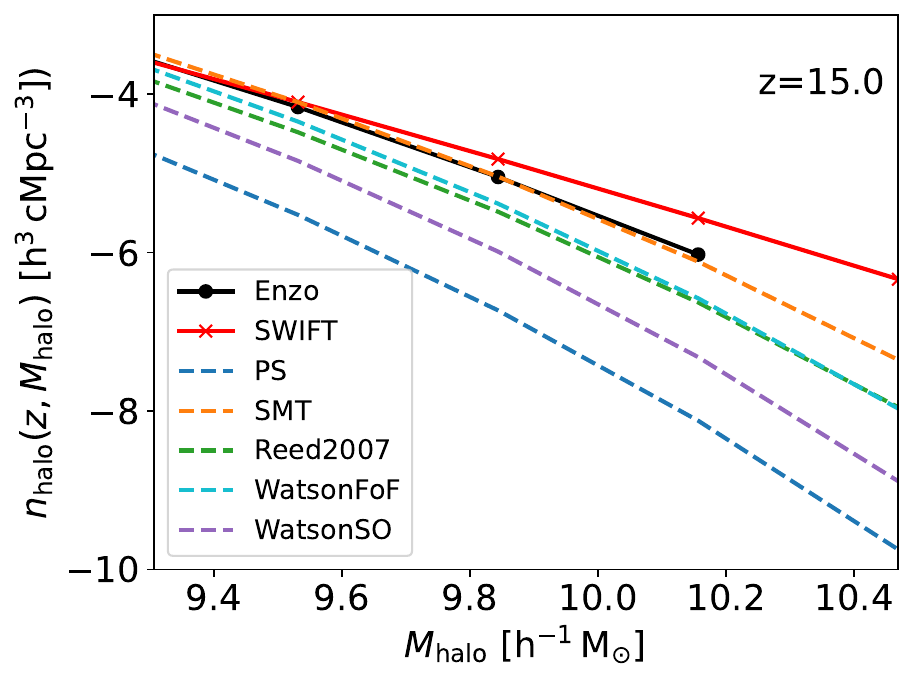}
    \hfill
        \includegraphics[width=0.4\textwidth]{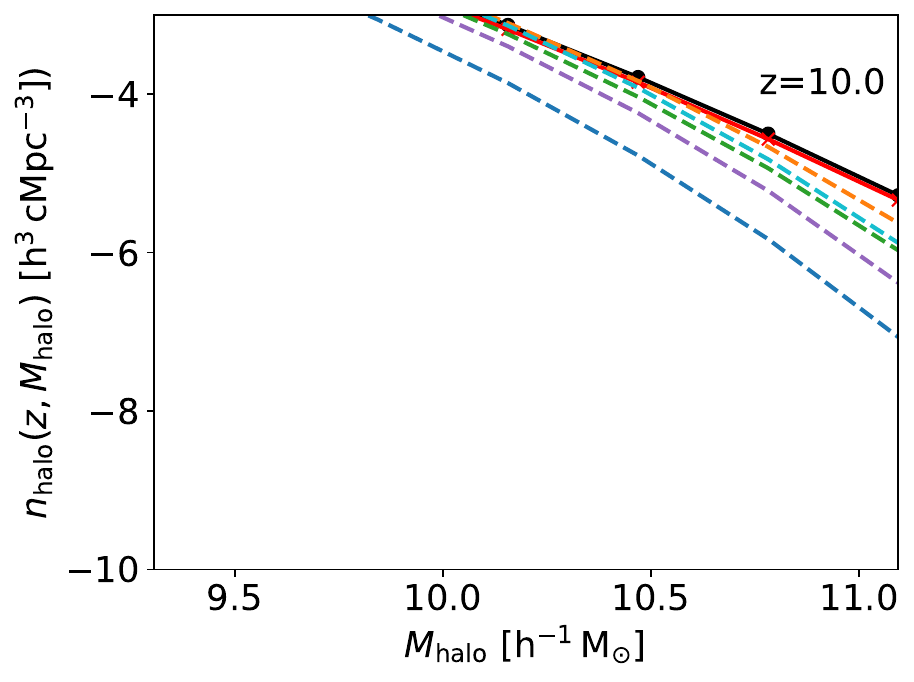}
        
\caption{A zoom-in onto the black rectangles identified in Figure \ref{fig:FOF} (with halo masses and number densities on a logarithmic scale). These mass ranges represent the most massive halos accessible via numerical simulation at these redshifts. At this higher level of detail we see some discrepancy between the numerical halo mass functions of \enzo and \swift and their (semi-)analytical counterparts, particularly at $z=15.0$. The residuals for each of the lines are shown in Figure \ref{fig:EnzoVFOF} and Figure \ref{fig:SWIFTVFOF}.}
\label{fig:FOFZoom}
\end{figure}
\subsubsection{Press-Schechter Theory}
\noindent In \citet{press_formation_1974} (PS), a spherical collapse model is assumed for dark matter halos and the probability of reaching a threshold density field value follows a Gaussian distribution. Thus the  fitting function for PS is given as:

\begin{equation} \label{eqn:PS_f}
    f_{\text{PS}}(\sigma) = \sqrt{\frac{2}{\pi}}\frac{\delta_{\text{c}}}{\sigma}\exp\Big(-\frac{\delta_{\text{c}}^{2}}{2\sigma^{2}}\Big),
\end{equation}

where $\delta_{\text{c}} \approx 1.686$ is the critical overdensity required for a region to spherically collapse into a dark matter halo.
A limitation of the PS fitting function is its tendency to overestimate the number of lower-mass halos and underestimate the number of higher-mass halos compared to N-body simulations (\citet{Lacey_1994}, \citet{Sheth_1999}). Since this seminal paper, many other fitting functions have been developed that aim to correct for this discrepancy as well as attempting to add additional sophistication to the modelling. In this work, we explore four other halo mass fitting functions in addition to PS.

\subsubsection{Sheth, Mo \& Tormen and Beyond}
\noindent The \citet{sheth_ellipsoidal_2001} (SMT) halo mass function has a similar form to PS but assumes ellipsoidal collapse and is given as:

\begin{equation} \label{eqn:SMT_f}
    f_{\text{SMT}}(\sigma) = A\sqrt{\frac{2a}{\pi}}\left[1 + \left(\frac{\sigma^{2}}{a\delta_{c}^{2}}\right)^{p} \right]\frac{\delta_{\text{c}}}{\sigma}\exp\left(-\frac{a\delta_{\text{c}}^{2}}{2\sigma^{2}}\right),
\end{equation}

where $A=0.3222$, $a=0.707$ and $p=0.3$.
We also test fitting functions developed by \citet{reed_halo_2007} (Reed07) and \citet{Watson_2013} (WatsonFoF refers to a fit using the Friends-of-Friends halo finder, WatsonSO refers to a fit using the Spherical Overdensity halo finder).
The PS and SMT fits are widely-used and based on an analytical formalism, with no restriction on redshift or halo mass range. The Reed07, WatsonFoF and WatsonSO fits are based on simulation results, all with a redshift range of $z \, \in (0, 30)$ and no explicit restriction on halo mass (see Table \ref{tab:HMFs}). While many other halo mass fitting functions exist, we choose these fits as they were all calibrated across the redshift range of interest to us (i.e. $z > 10$). Other fitting functions are typically calibrated for lower redshifts. 

We use the Python package \texttt{hmf} to compute halo number densities derived from the HMFs described above.

\begin{table}
    \centering
    \begin{tabular}{|l|l|l|}
    \hline
         Fitting function & $z$ range & Reference\\
         \hline \hline
         PS & No limit & \citet{press_formation_1974} \\
         SMT &  No limit & \citet{sheth_ellipsoidal_2001} \\
         Reed2007 &  0 - 30 & \citet{reed_halo_2007} \\
         WatsonFoF &  0 - 30 & \citet{Watson_2013} \\
         WatsonSO &  0 - 30 & \citet{Watson_2013} \\
         \hline
    \end{tabular}
    \caption{The fitted halo mass functions we use to compare to HMFs derived from numerical simulations. }
    \label{tab:HMFs}
\end{table}

\section{Results} \label{Sec:Results}
As previously stated, the goal of this paper is to compare halo number densities derived from simulation results with those derived from popular fits, and to test how well the fits can predict JWST halo mass abundances compared to simulations \citep[e.g.][]{Boylan-Kolchin_2023}.
\begin{figure}[h]
    \centering
         \includegraphics[width=0.4\textwidth]{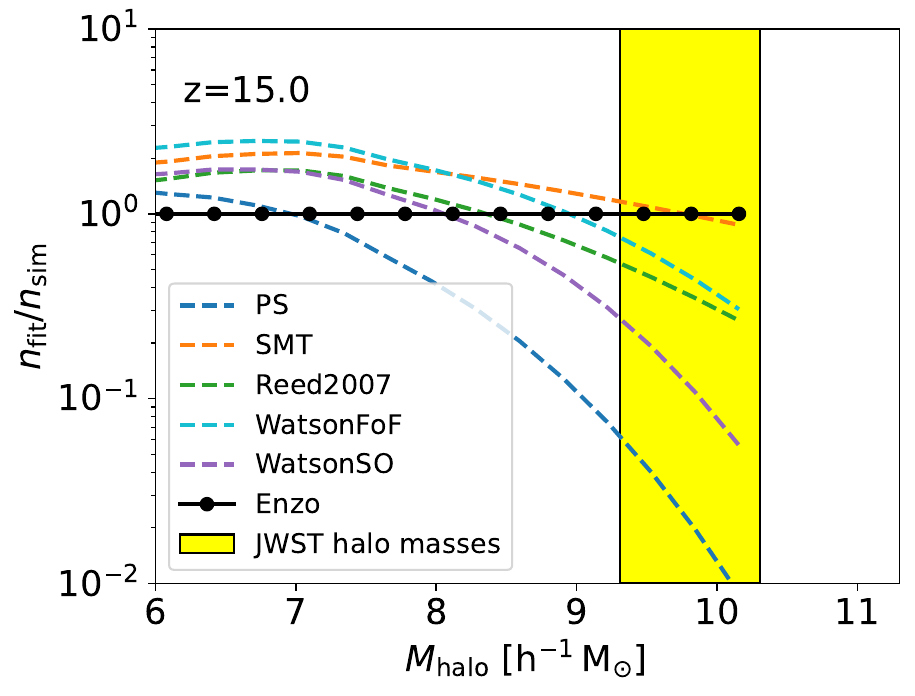}
    \hfill
        \includegraphics[width=0.4\textwidth]{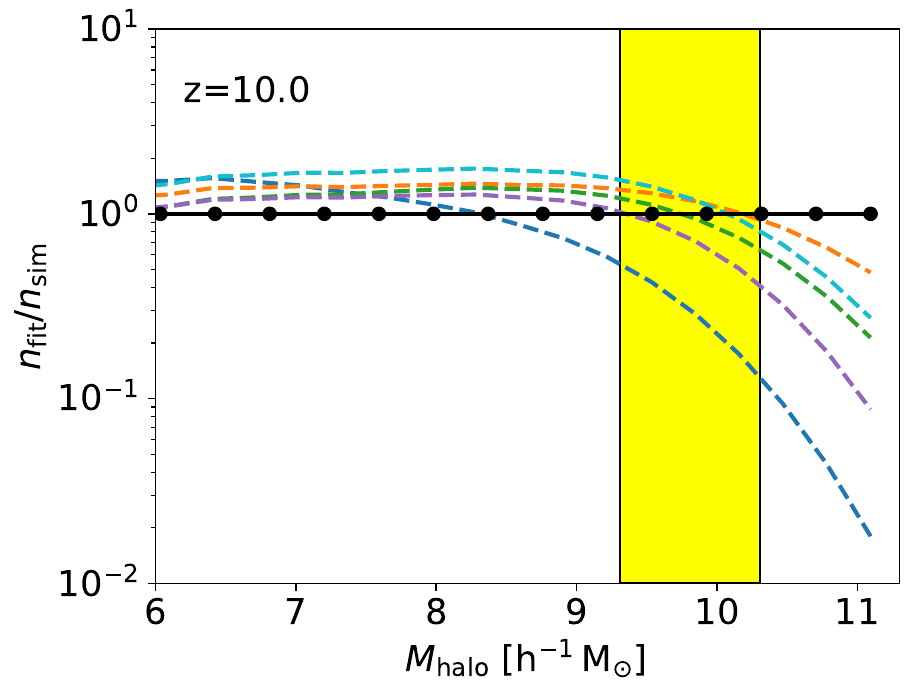}
    \caption{The ratios of the fitted halo number densities to the numerical halo number densities, derived from \enzo simulations and using the FOF halo finder (with halo masses on a logarithmic scale). Differences between \enzo and the (semi-)analytical HMFs are typically less than a factor of two at $z=10.0$ increasing to a factor of 5 at $z=15.0$ (excluding the PS fit).}
    \label{fig:EnzoVFOF}
\end{figure}

\begin{figure}[h]
    \centering

        \includegraphics[width=0.4\textwidth]{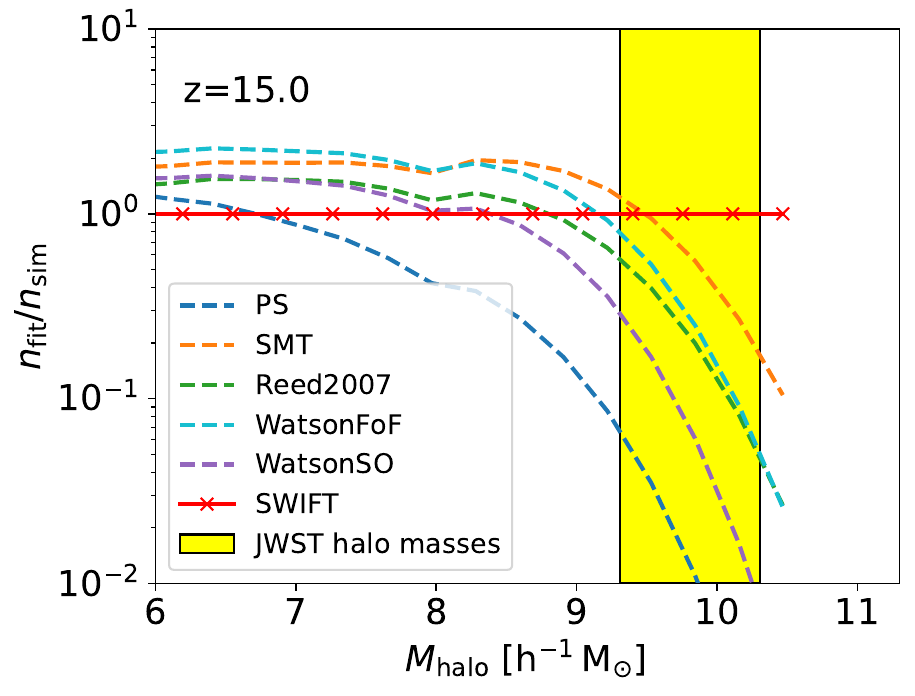}
    \hfill
        \includegraphics[width=0.4\textwidth]{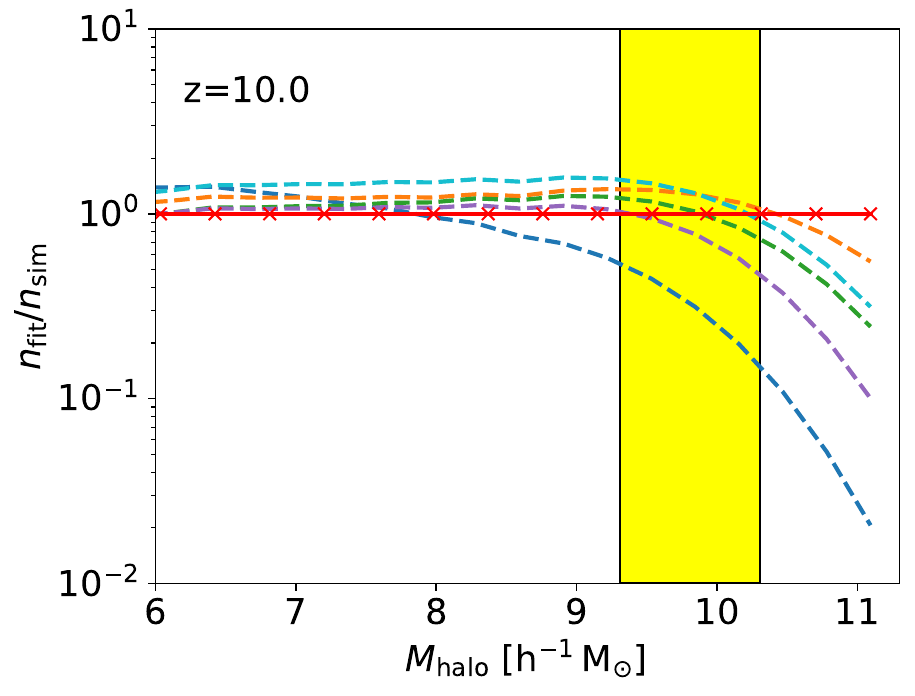}
     
    \caption{The ratios of the fitted halo number densities to the numerical halo number densities, derived from \swift simulations and using the FOF halo finder (with halo masses on a logarithmic scale). Similar to the \enzo result, \swift shows excellent agreement with the (semi-) analytic HMFs, all fits except PS agreeing within a factor of 2 for all but the highest masses at $z=10.0$. There is a greater discrepancy present at $z=15.0$, with some fits underestimating the numerical results by a factor of 100 at the high-mass range.}
    \label{fig:SWIFTVFOF}
\end{figure}

In Figure \ref{fig:FOF}, we compare the HMFs from \enzo and \swift data (dashed lines) with the chosen fits (solid lines) for $z=20.0$, $z=15.0$ and $z=10.0$ using the FOF halo finder. 
The black rectangles indicate the high mass range depicted in more detail in Figure \ref{fig:FOFZoom}. The range of halo masses selected in the black rectangles is bounded by the approximate lowest halo masses observable by JWST at these epochs up to the most massive halo masses accessible by the numerical simulations at that redshift. This is only possible for our outputs at $z=10.0$ and $z=15.0$ as halo masses large enough to be observed (except perhaps via extreme lensing) by JWST are simply not formed by $z=20.0$ in a $\Lambda$CDM universe. Hence, we focus our analysis on the $z=15.0$ and $z=10.0$ outputs in particular.
Overlaid onto Figures \ref{fig:FOF} to \ref{fig:SWIFTVFOF} are yellow rectangles which act as a visual aid, depicting an estimate for the range of halo masses that host recently-observed JWST galaxies and galaxy candidates.
We choose sources with $10.0 < z < 15.0$ \footnote{We use photometric redshift only if spectroscopic redshift is not yet available.} and with an estimate of $M_{*}$ or $M_{\text{halo}}$ available. These sources include GN-z11 \citep{Scholtz_2024}, Maisie's Galaxy \citep{Haro_2023}, GS-z14 \citep{Helton_2024} and others \citep{Chakraborty_2024}.
If there exists no estimate of $M_{\text{halo}}$ yet for a given source, we estimate it using the following equation: 

\begin{equation}
    M_{\text{halo}}(z, M_{*}) = \frac{\Omega_{m}(z)}{\Omega_{b}(z)}\frac{M_{*}}{\epsilon_{*}},
\end{equation}

where we set the star formation efficiency parameter $\epsilon_{*} = 0.1$. 
The width of the yellow rectangle is determined by the minimum and maximum halo masses in our sample and does not vary with redshift due to the small number of sources.
In Figures \ref{fig:EnzoVFOF} and \ref{fig:SWIFTVFOF}, we take the ratios of the semi-analytic halo number densities against the \enzo and \swift outputs respectively to quantify the overall disagreements at $z=10.0$ and $z=15.0$.

Over the full mass range, for both simulation suites and halo finders, we find the fits considered agree well across the entire spectrum. There is, over several dexes in mass and over the redshift range between $z=10.0$ and $z=20.0$, excellent agreement between all of the HMFs. Additionally, the fits improve as redshift decreases - not just between the numerical results and the (semi-)analytic results but even among the (semi-)analytic fits themselves.

At the lower mass range ($10^{6}$ to $10^{7}$ h$^{-1}$ $\text{M}_{\odot}$), the semi-analytic fits overestimate the halo number densities by up to a factor of $\approx$ 2 compared to either \enzo or \swift data.

At the mid-range masses (10$^{7}$ to 10$^{9}$ h$^{-1}$ M$_{\odot}$) at $z=10.0$, all but the PS fit overestimate the numerical simulations by less than a factor of 2. This is best illustrated by the bottom panels of Figures \ref{fig:EnzoVFOF} and \ref{fig:SWIFTVFOF}.
For the same mass range at $z=15.0$, there is a greater discrepancy with all but the SMT and WatsonFOF fits underestimating the halo number densities by up to a factor of $\approx$ 5 compared to numerical simulations (see the top panels of Figures \ref{fig:EnzoVFOF} and \ref{fig:SWIFTVFOF}).

In the JWST mass range, we see the greatest discrepancies. At $z=15.0$, the PS semi-analytic fit underestimates the \enzo halo number density by a factor of $\approx$ 100 (see Figure \ref{fig:EnzoVFOF} (top panel)). The other fits agree within a factor of $\approx$ 50 (WatsonSO) in the worst case with the SMT fit providing the closest match. We see a similar story at $z=10.0$ with a less extreme discrepancy (PS underestimating by a factor of $\approx$ 10, the other fits agreeing within a factor of $\approx$ 2 (see Figure \ref{fig:EnzoVFOF} (bottom panel))). This is consistent with the results and conclusions of \cite{Yung_2024}.\\
For the \swift simulations at $z=15.0$, we see a strong deviation towards higher masses which was not present for the \enzo simulations (see Figure \ref{fig:SWIFTVFOF} (top panel)). Both PS and WatsonSO underestimate the numerical halo number densities by a factor of $\approx$ 100, and SMT agrees best within a factor of $\approx$ 10. It appears here that the \enzo runs are better able to resolve halos at earlier times perhaps due to the inherent refinement strategy. A detailed analysis of the difference between the numerical codes is outside the scope of this paper and comparisons in this direction have been undertaken in the past \citep[e.g.][]{OShea_2005, Regan_2007, Hayward_2014}. 
At $z=10.0$ the agreement between the numerical (\texttt{SWIFT}) results and the analytic fits remains excellent (see Figure \ref{fig:SWIFTVFOF} (bottom panel)). The PS fit underestimates the numerical results by a factor of $\approx$ 10, with the other fits agreeing within a factor of 2.
In Figure \ref{fig:FOFZoom} (top panel), we see a discrepancy between the \enzo and \swift results in the high-mass range at $z=15.0$. Previous work \citep[e.g.][]{Warren_et_al2006,More_at_al_2011} has shown that the FOF mass is sensitive to mass resolution and the presence of substructure, which will be strongly dependent on redshift and which is difficult to correct for in general. It will also depend on local clustering, which can lead to distinct structures being linked by bridges of particles. This is the most likely explanation for the variance between \enzo and \swift at $z=15.0$. Note that this variance is absent from the HOP profiles (see Figure \ref{fig:HOPZoom}).

In summary the agreement between the numerical N-body solvers and the (semi-)analytic fits is excellent. In particular, the SMT fitting function gives excellent agreement with \texttt{Enzo}, with deviations of at most approximately a factor of 2 at $z=15.0$ (for low halo masses which are anyway currently unobservable) and converging to much less than a factor of two within the JWST window at $z=10.0$. For \swift, the agreement with SMT is equally excellent at $z=10.0$ but deviates somewhat for the higher halo masses at $z=15.0$.  Using such (semi-)analytic fits, then, is unlikely to be a dominant source of error when testing high redshift observations against $\Lambda$CDM models.

\section{Discussion} \label{Sec:Discussion}
\noindent The recent explosion of data from the high-$z$ Universe, particularly beyond $z=10$, by JWST has led to a number of claims that the data is in tension with our galaxy formation and cosmological models 
\citep[e.g.][]{Boylan-Kolchin_2023, ArrabalHaro_2023, Yung_2024, Finkelstein_2024}. The Universe beyond $z=10$ however is likely to be significantly different to the later and present-day Universe. At $z \gtrsim 10$, galaxies are still in their infancy, with the most massive galaxies at those epochs having stellar masses less than $10^{10} \ \msolar$  (these would be classified as dwarfs in the present-day Universe). Moreover, there is strong evidence that the astrophysical processes at play at $z=10$ are sufficiently different to those of the present-day Universe and that they make significant alterations to the galaxy properties. This is particularly evident in galaxies like GN-z11 which is thought to harbour a massive black hole at its centre \citep{Maiolino_2023}, moreover this galaxy contains species abundances which are difficult to explain through standard processes \citep[e.g.][]{Bunker_2023, Cameron_2023, Charbonnel_2023, Nandal_2024}. This peculiarity and lack-of-understanding is not unique to GN-z11 with a number of galaxies displaying properties which has evoked confusion within the community \citep[e.g.][]{Maiolino_2024Xray}. The most luminous galaxies observed by JWST remain in tension with state-of-the-art cosmological simulations (e.g. \cite{Keller_2023}) with simulations struggling to model their extreme brightness at very early times. The reasons behind this are currently unknown but a greater emphasis on processes specific to the early Universe such as population III star formation and early black hole formation (e.g. \cite{McCaffrey_2023}) may offer a pathway forward. It has also been argued that this tension may be resolved with the use of high-resolution simulations targeted at the high-redshift universe (see \cite{McCaffrey_2023}). \\
\indent A key part of making progress in understanding the high-$z$ Universe is therefore to identify sources of 
systematic error in our models at high-$z$. 
Some sources of uncertainity in counting galaxies at such redshifts include cosmic variance (especially significant at the distances considered), error in stellar mass estimation and the presence of backsplash halos (halos that have lost some dark matter from the host halo, giving the impression of a higher baryon-dark matter ratio). These uncertainties have been explored thoroughly by \citet{Chen_2023}.
In this paper we focus on exploring differences in fitting functions to the universal HMF and how they compare against direct N-body simulations at $z \geq 10$. In particular, we 
compare a wide range of fitting functions in use in the literature (see Table \ref{tab:HMFs}) against the adaptive mesh refinement code \enzo and against the N-body SPH code \texttt{SWIFT}. \\
\indent We find that, for both \enzo and \swift, the match against the HMF (semi-)analytic functions is excellent, with many fits agreeing with numerical results within a factor of 2 for low and mid-range halo masses. The match against the standard \cite{press_formation_1974} formalism is less accurate with deviations of up to an order of magnitude (at $z=10.0$ and $z=15.0$). Similarly, when comparing against standard fitting formula (see Table \ref{tab:HMFs}) we again see good agreement with deviations typically within an order of magnitude up to $z=15.0$ inside the window in which JWST can approximately observe high-$z$ galaxies. \\
\indent We caution that the spatial resolution employed, controlled via the softening parameter for \swift and via the level of maximum refinement with \enzo, is set relatively high for our simulations. For example, we 
use a softening length set to approximately the mean interparticle spacing divided by 25. This is slightly lower than the resolution we evolve the \enzo simulations with. This level of gravity resolution, specifically the softening length and spatial resolution here, is likely to be significantly higher than that used for typical galaxy formation simulations designed to run to $z \sim 0$ \citep[e.g.][]{Power_2003, Power_2016,Zhang_2019}. It is also worth noting that the gravitational collapse of dark matter halos is sensitive to the large-scale gravitational field, which in a numerical simulation depends on the size of the simulation volume. Previous work \citep[e.g.][]{Power_and_Knebe_2006} has highlighted how the mass function is sensitive to simulation volume, with a deficit of halos of a given mass at high masses. This effect is most pronounced in studies of the mass function at high redshifts, where the necessary high mass resolutions and large simulation volumes make these simulations particularly challenging. We therefore caution the reader that matching the results from numerical simulations designed primarily for large-scale investigations may struggle to calculate the correct halo properties and abundances at $z \gtrsim 10$ \citep[see e.g.][]{Keller_2023} unless the simulations are truly focused on high-$z$ study \cite[see e.g.][]{McCaffrey_2023}. \\
\indent Nonetheless, overall we find excellent agreement between N-body simulations and both (semi-)analytic and fitting functions to HMFs, consistent with the results from \cite{Yung_2024}, and that these functions are unlikely to lead to large errors in our modelling of high-$z$ host halos when appropriately modelled. 

\section*{Data availability}
The analysis code used to produce the figures shown here is available upon request.

\section*{Acknowledgements}
\noindent  JR acknowledges support from the Royal Society and Science Foundation Ireland under grant number URF\textbackslash R1\textbackslash 191132 and support from the Irish Research Council Laureate programme under grant number IRCLA/2022/1165. CP acknowledges the support of the Australian Research Council Centre of Excellence for All Sky Astrophysics in 3 Dimensions (ASTRO 3D), through project number CE170100013. HOB thanks JR's research group for many helpful discussions and insights: John Brennan, Lewis R. Prole, Saoirse Ward, Joe McCaffrey, Pelle van de Bor and Daxal Mehta. The authors wish to acknowledge the Irish Centre for High‑End Computing (ICHEC) for the provision of computational facilities and support via the Meluxina facility, hosted by LuxProvide (Luxembourg). We wish to thank the EuroHPC Joint Undertaking for awarding this project access to the EuroHPC supercomputer Karolina, hosted by IT4I (Czech Republic) through a EuroHPC Regular Access call. We also thank the anonymous referee for a constructive and insightful report.

\bibliographystyle{mn2e}
\bibliography{references}

\appendix
\counterwithin{figure}{section}

\section{Alternative Halo Finding techniques}
\begin{figure}[h]
    \centering
        \includegraphics[width=0.5\textwidth]{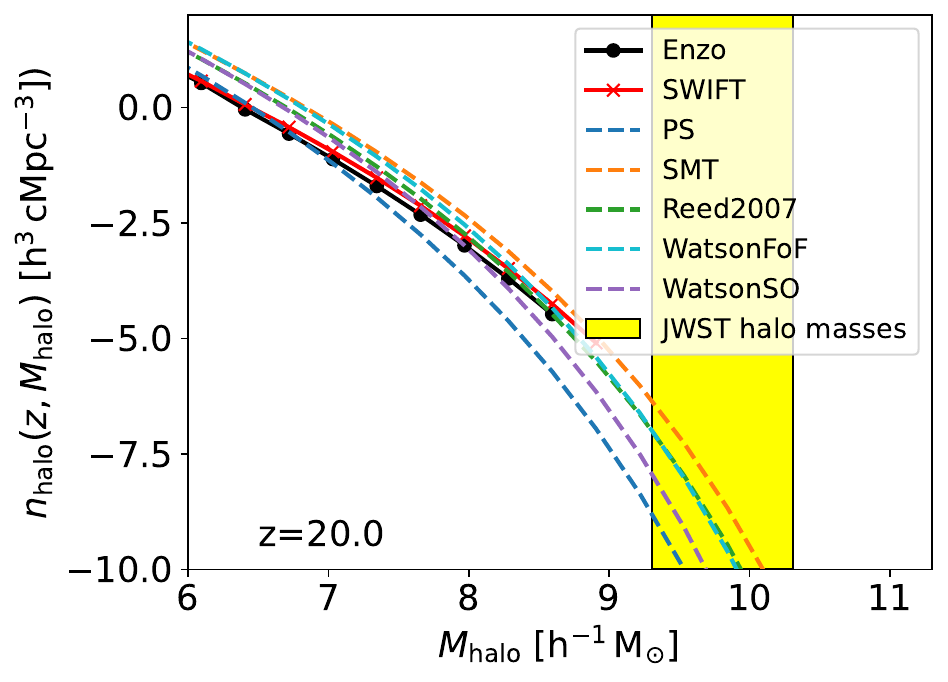}
    \hfill
        \includegraphics[width=0.5\textwidth]{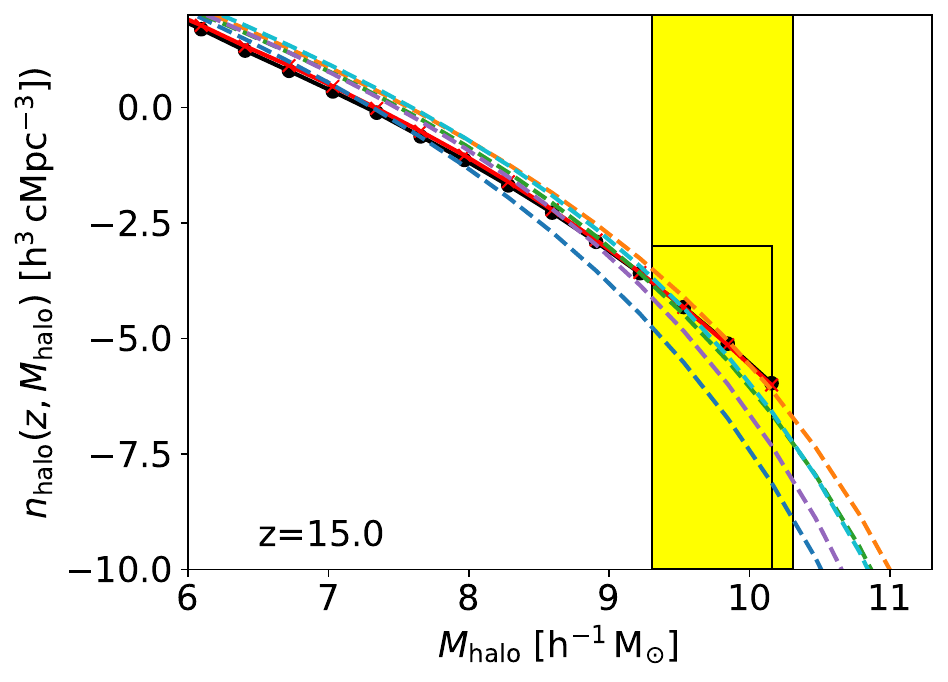}
    \hfill
        \includegraphics[width=0.5\textwidth]{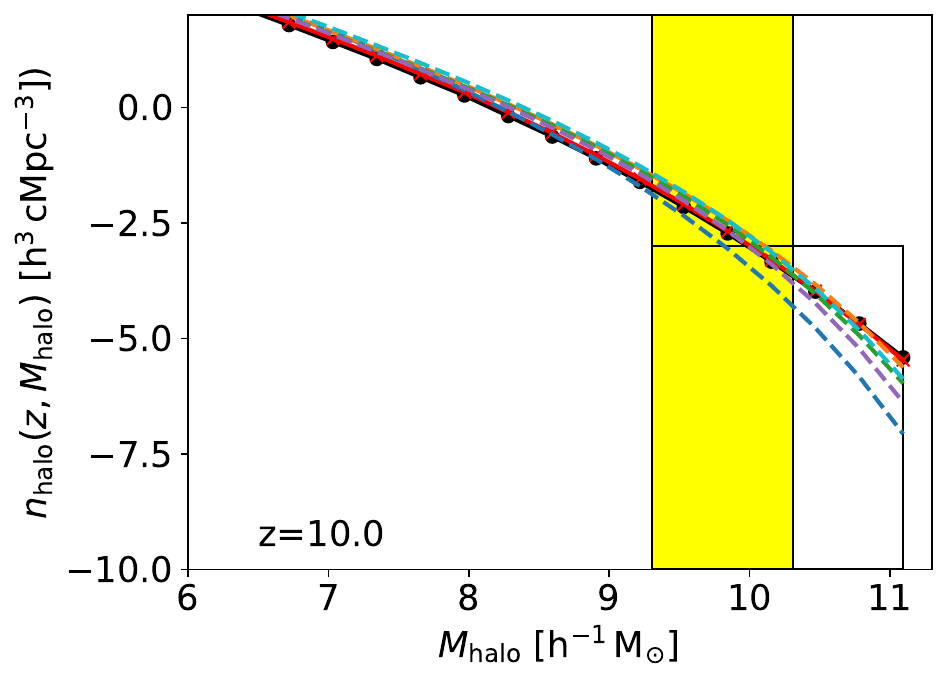}
\caption{Comparing the \enzo and \swift halo number densities contrasted with 5 halo number densities derived from fits at $z=20.0$ (upper panel), $z=15.0$ (centre panel) and $z=10.0$ (bottom panel) using the HOP halo finder (with halo masses and number densities on a logarithmic scale).}
\label{fig:HOP}
\end{figure}

\begin{figure}[h]
    \centering
        \includegraphics[width=0.45\textwidth]{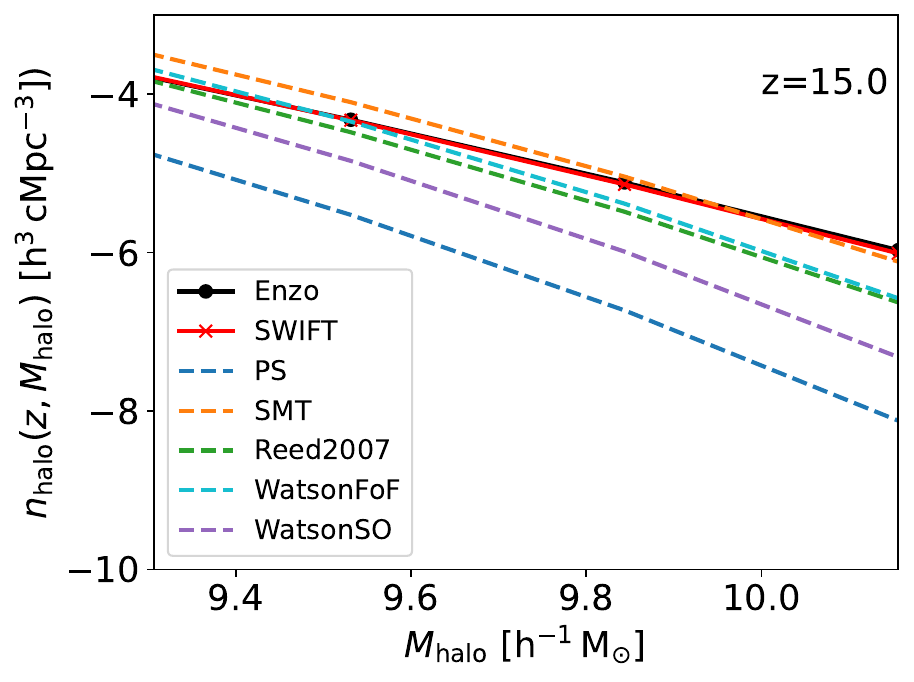}
    \hfill
        \includegraphics[width=0.45\textwidth]{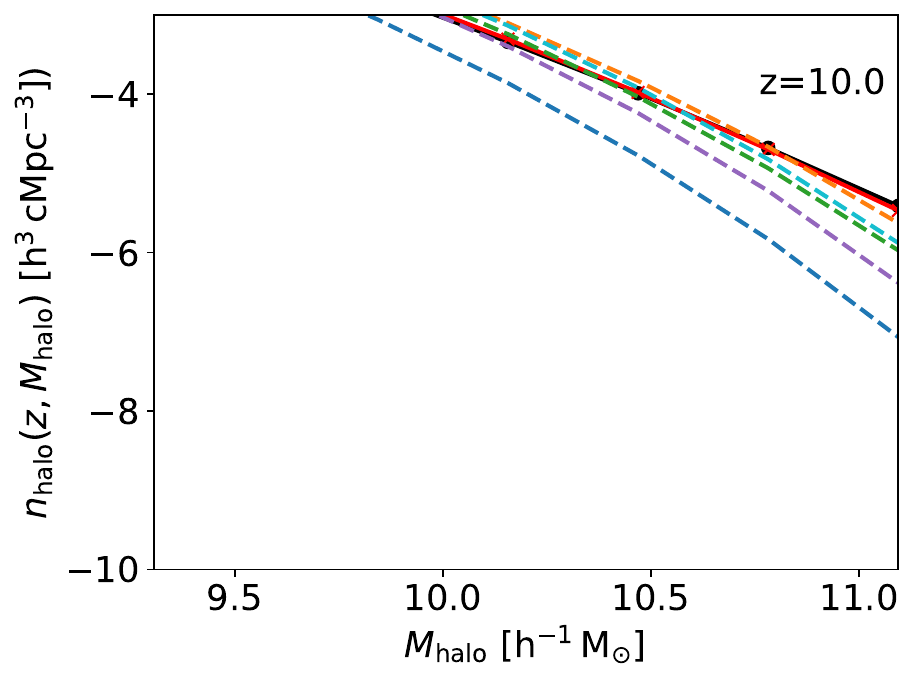}
     
\caption{High-mass sections of Figure \ref{fig:HOP} (centre and bottom panels) shown in more detail.}
\label{fig:HOPZoom}
\end{figure}

\begin{figure}[h]
    \centering
        \includegraphics[width=0.45\textwidth]{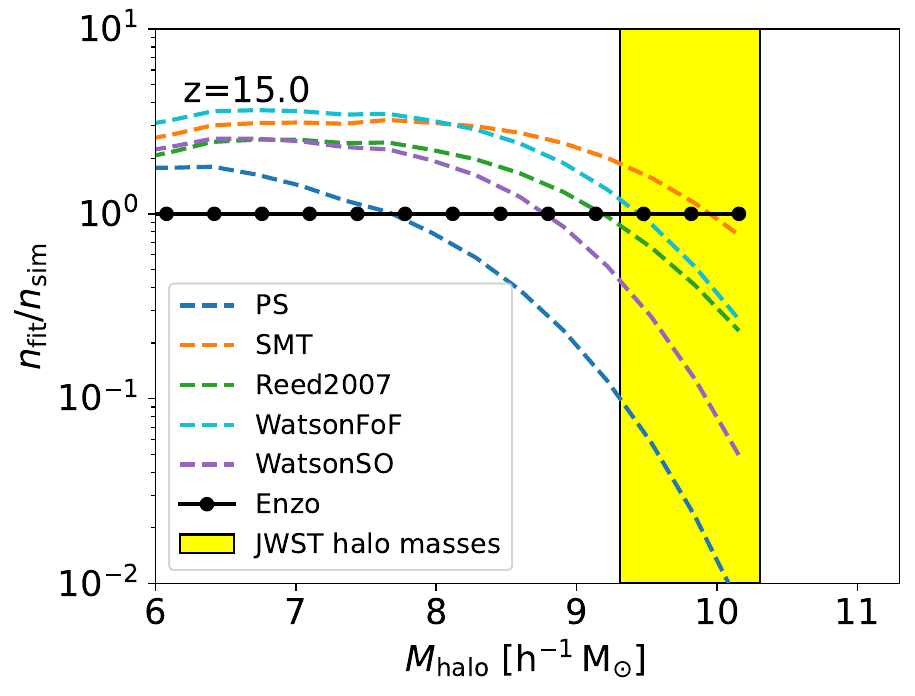}
    \hfill
        \includegraphics[width=0.45\textwidth]{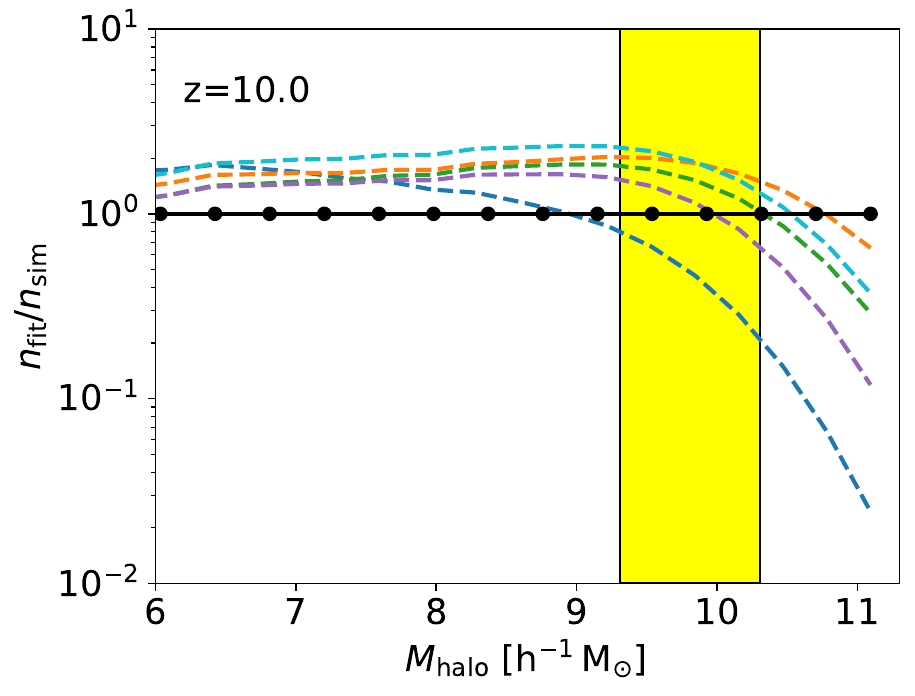}
    \caption{The ratios of the fitted halo number densities to the numerical halo number densities, derived from \enzo simulations and using the HOP halo finder (with halo masses on a logarithmic scale).}
    \label{fig:EnzoVHOP}
\end{figure}

\begin{figure}[h]
    \centering
        \includegraphics[width=0.45\textwidth]{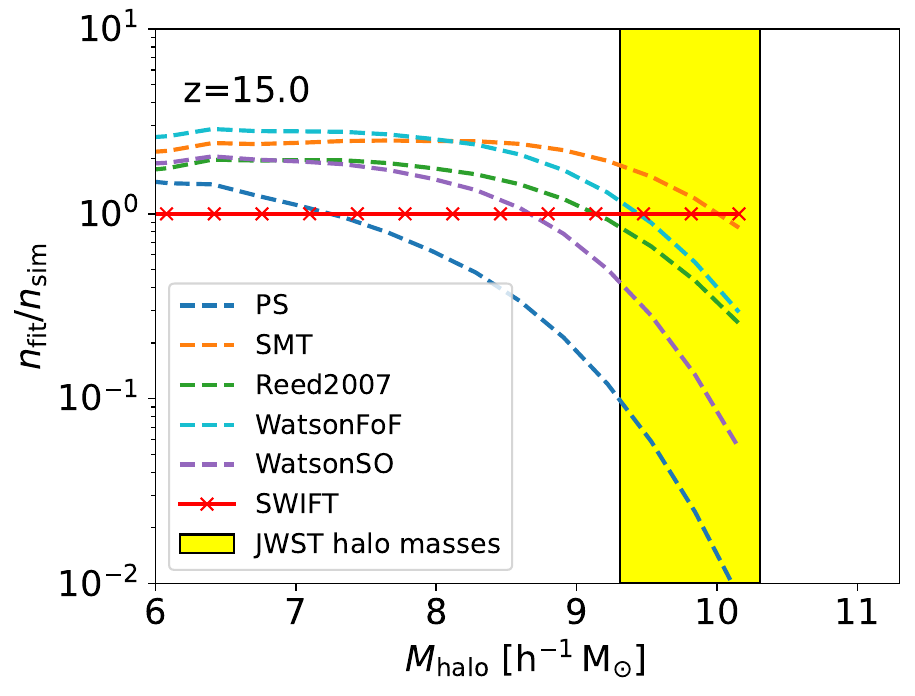}
    \hfill
        \includegraphics[width=0.45\textwidth]{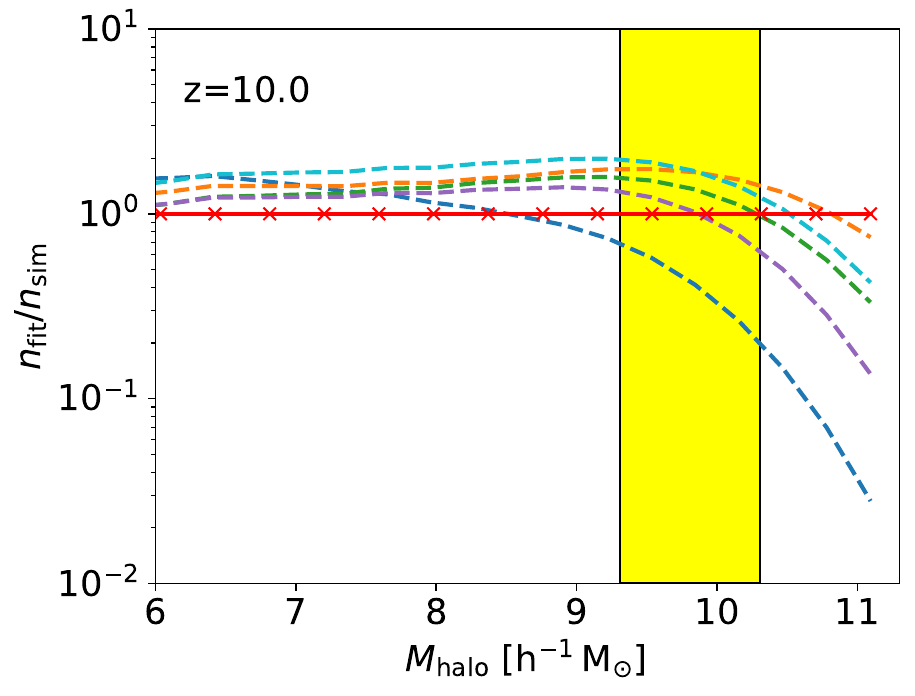}
     
    \caption{The ratios of the fitted halo number densities to the numerical halo number densities, derived from \swift simulations and using the HOP halo finder (with halo masses on a logarithmic scale).}
    \label{fig:SWIFTVHOP}
\end{figure}

In addition to the widely-used Friends-of-Friends halo finding algorithm, we also investigate how a different halo finder would impact our results. As discussed in \S \ref{Sec:Methodology} we also use the HOP halo finder (\citet{eisenstein_hop_1998}) in our analysis. 
In Figures \ref{fig:HOP} to \ref{fig:SWIFTVHOP}, we reproduce Figures \ref{fig:FOF} to \ref{fig:SWIFTVFOF} from the main text respectively but with the halo finding technique switched to HOP instead of FOF. 
Using the HOP halo finder we find near identical results to the FOF method and again we see approximately a factor of two difference between HOP and the analytic HMFs at both $z=10.0$ and $z=15.0$. The difference between HOP and the fitting functions varies up to an order of magnitude but similar to the FOF method the agreement is nonetheless still very good - particularly at $z=10.0$ where differences are typically less than a factor of two. We see negligible differences between FOF and HOP  - a finding supported by other research studies \citep{Knebe_2011}. \\ \\ \\

\end{document}